%% file: aerialBaseStations.tex
\def\editmode{0}
\def\reportmode{0}

\def\bibfilenames{WISENET}

\if\reportmode1
\documentclass[10pt,final,onecolumn]{IEEEtran}
\else
\documentclass[conference]{IEEEtran}
\IEEEoverridecommandlockouts
\fi

\if\editmode1  
\usepackage[backend=bibtex,style=alphabetic,sorting=debug]{biblatex}
\DeclareFieldFormat{labelalpha}{\thefield{entrykey}}
\DeclareFieldFormat{extraalpha}{}
\bibliography{\bibfilenames}
\newcommand{\cmt}[1]{\noindent\textcolor{darkgreen}{\underline{[#1]}}} 
\newcommand{\hc}[1]{\textcolor{blue}{#1}} 
\newenvironment{myitemize}{\begin{itemize}}{\end{itemize}}
\newcommand{\myitem}{\item}

\newcommand{\opt}[1]{\textcolor{darkgray}{#1}}
\newcommand{\nextv}[1]{}

\else
\usepackage{cite}
\newcommand{\cmt}[1]{} 
\newcommand{\hc}[1]{\textcolor{black}{#1}} 
\newenvironment{myitemize}{}{}
\newcommand{\myitem}{}

\newcommand{\opt}[1]{\textcolor{black}{#1}}
\newcommand{\nextv}[1]{}
\fi 

\input{include}

\begin{document}

\title{Non-cooperative Aerial Base Station \\Placement via Stochastic Optimization}

\if\reportmode1
  \author{Daniel Romero\\[.5cm]Technical Report v. \today}
\else
\author{Daniel Romero$^1$ and Geert Leus$^2$\\[1mm]
$^1$ Dept. of Information and Communication Technology,
University of Agder, Norway, daniel.romero@uia.no\\
$^2$ Faculty of EE, Mathematics and Computer Sci., TU Delft, The Netherlands, g.j.t.leus@tudelft.nl
\thanks{The work in this paper was supported in part by the Indo-Norwegian
  program of the Research Council of Norway under project LUCAT and by the KAUST-MIT-TUD-Caltech consortium grant OSR-2015-Sensors-2700 Ext. 2018.
 }
}
\fi

\newcommand{\acom}[1]{\textcolor{red}{\textbf{[#1]}}}

\maketitle
\begin{abstract}
  Autonomous unmanned aerial vehicles (UAVs) with on-board base
  station equipment can potentially provide connectivity in areas
  where the terrestrial infrastructure is overloaded, damaged, or
  absent. Use cases comprise emergency response, wildfire suppression,
  surveillance, and cellular communications in crowded events to name
  a few. A central problem to enable this technology is to place such
  aerial base stations (AirBSs) in locations that approximately
  optimize the relevant communication metrics. To alleviate the
  limitations of existing algorithms, which require intensive and
  reliable communications among AirBSs or between the AirBSs and a
  central controller, this paper leverages stochastic optimization and
  machine learning techniques to put forth an adaptive and
  decentralized algorithm for AirBS placement without inter-AirBS
  cooperation or communication. The approach relies on a smart design
  of the network utility function and on a stochastic gradient ascent
  iteration that can be evaluated with information available in
  practical scenarios. To complement the  theoretical
  convergence properties, a simulation study
  corroborates the effectiveness of the proposed scheme.



\end{abstract}

\if\reportmode0
\begin{keywords}
Aerial communications, unmanned aerial vehicles, aerial base stations,
stochastic optimization, stochastic 
gradient, autonomous aerial vehicles, drones.
\end{keywords}
\fi

\renewcommand{\transpose}{^\top}
\renewcommand{\expected}{\mathbb{E}}

\newcommand{\bstext}{{\hc{\text{BS}}}}
\newcommand{\mutext}{{\hc{\text{MU}}}}

\newcommand{\bslocvec}{\hc{\bm l}}
\newcommand{\bslocset}{\hc{\mathcal{L}}}
\newcommand{\bsflatlocvec}{\hc{\bbm l}}

\newcommand{\bsnot}[1]{_{#1}}
\newcommand{\bsmunot}[2]{_{#1,#2}}
\newcommand{\bsind}{{\hc{b}}}
\newcommand{\bsnum}{{\hc{B}}}
\newcommand{\txpower}{\hc{P}}
\newcommand{\height}{\hc{h}}

\newcommand{\propconst}{\hc{K}}

\newcommand{\mulocvec}{\hc{\bm x}}
\newcommand{\muflatlocvec}{\hc{\bbm x}}
\newcommand{\munum}{{\hc{M}}}
\newcommand{\muind}{{\hc{m}}}
\newcommand{\auxmuind}{{\hc{\tilde m}}}
\newcommand{\munot}[1]{_{#1}}

\newcommand{\packnum}{{\hc{S}}}
\newcommand{\packind}{{\hc{s}}}
\newcommand{\packnot}[1]{[#1]}

\newcommand{\chgain}{\hc{g}}
\newcommand{\rxpower}{\hc{p}}
\newcommand{\minrxpower}{\hc{p_\text{min}}}
\newcommand{\maxrxpower}{\hc{\infty}}

\newcommand{\probpack}{\hc{\pi}} 

\newcommand{\itnot}[1]{[#1]}
\newcommand{\itnum}{{\hc{I}}}
\newcommand{\itind}{{\hc{i}}}

\newcommand{\stepsize}{{\hc{\eta}}}
\newcommand{\noisepower}{{\hc{N_0}}}
\newcommand{\rate}{{\hc{R}}}
\newcommand{\minrate}{{\hc{R_\text{min}}}}
\newcommand{\minpower}{{\hc{p_\text{min}}}}
\newcommand{\maxpower}{{\hc{p}}^\text{\hc{max}}}
\newcommand{\sumpower}{{\hc{p}}^\text{\hc{sum}}}
\newcommand{\unitstep}{{\hc{u}}}

\newcommand{\minibatchnum}{{\hc{Q}}}
\newcommand{\logsumexpfun}{{\hc{\phi}}}

\section{Introduction}
\label{sec:intro}

\cmt{Motivation}
\begin{myitemize}
  \myitem\cmt{motivation UAVs}A widespread recognition among the
  general public prescribes that unmanned aerial vehicles (UAVs) will sooner
  or later serve for a number of applications with a transformative
  impact on human life. 
\myitem\cmt{motivation AirBSs}For example, a
collection of base stations 
on board autonomous UAVs can be deployed 
to provide data connectivity in areas where the  communication
infrastructure is overloaded, damaged, or absent; see
e.g.~\cite{zeng2019accessing}. Such a technology may benefit
\begin{myitemize}%
\myitem wildfire suppression,
\myitem search and rescue operations,
\myitem communications in crowded demonstrations  or sport events (even out of
cities such as bicycle races), 
\myitem emergency response, and even  natural disaster management.  
\end{myitemize}%
\myitem\cmt{motivation placement}This application requires algorithms
that allow such aerial base stations (AirBSs) to adopt positions in 3D
space that approximately optimizes quality of service (QoS) to the
mobile users (MUs)  without 
compromising operational safety.
\end{myitemize}%

\cmt{Literature}%
\begin{myitemize}%
  \myitem\cmt{one UAV}The problem of 
    \emph{AirBSs placement}  has been addressed for a
  single AirBS e.g. in~\cite{boryaliniz2016placement,chen2017map,han2009manet,lee2011climbing}.
\myitem\cmt{multiple UAVs}To accommodate multiple AirBSs, 
\begin{myitemize}%
\myitem\cmt{Centralized schemes}centralized
schemes have been proposed
\begin{myitemize}%
\myitem\cmt{list}in
\cite{liu2019deployment,kim2018topology,mozaffari2016coverage,wang2018adaptive},
\myitem\cmt{limitations}but they require a
central controller that receives all system information in real time
and instructs the AirBSs on how to navigate, which may be problematic
in practice due to the unreliable nature of wireless communications.
\end{myitemize}%
\end{myitemize}%
\myitem\cmt{decentralized}To alleviate this limitation, decentralized
schemes
\begin{myitemize}%
\myitem\cmt{need for communication}that require communication and
coordination only among 
neighboring AirBSs as well as between AirBSs and MUs have been developed 
\begin{myitemize}%
\myitem\cmt{list}in~\cite{lee2010decentralized,park2018formation}.
\myitem\cmt{limitations}However, failures of inter-AirBS
communications or  malfunctioning  AirBSs may compromise
 safety. Besides, only updating the neighborhood of each AirBS
in real time is error-prone and entails a considerable overhead. 
\end{myitemize}%
\end{myitemize}%
\myitem\cmt{Other}Other works are omitted due to space limitations;
see e.g. references
in~\cite{zeng2019accessing}.  \myitem\cmt{Summary}To the best
of our knowledge, all existing schemes for multiple-AirBS placement
are either centralized, or require intensive communication and coordination among
AirBSs, or are not amenable to adaptive implementations, as necessary
since  practical scenarios are subject to constant change.
\end{myitemize}%

\begin{figure}[t]
 \centering
 \includegraphics[width=0.45\textwidth]{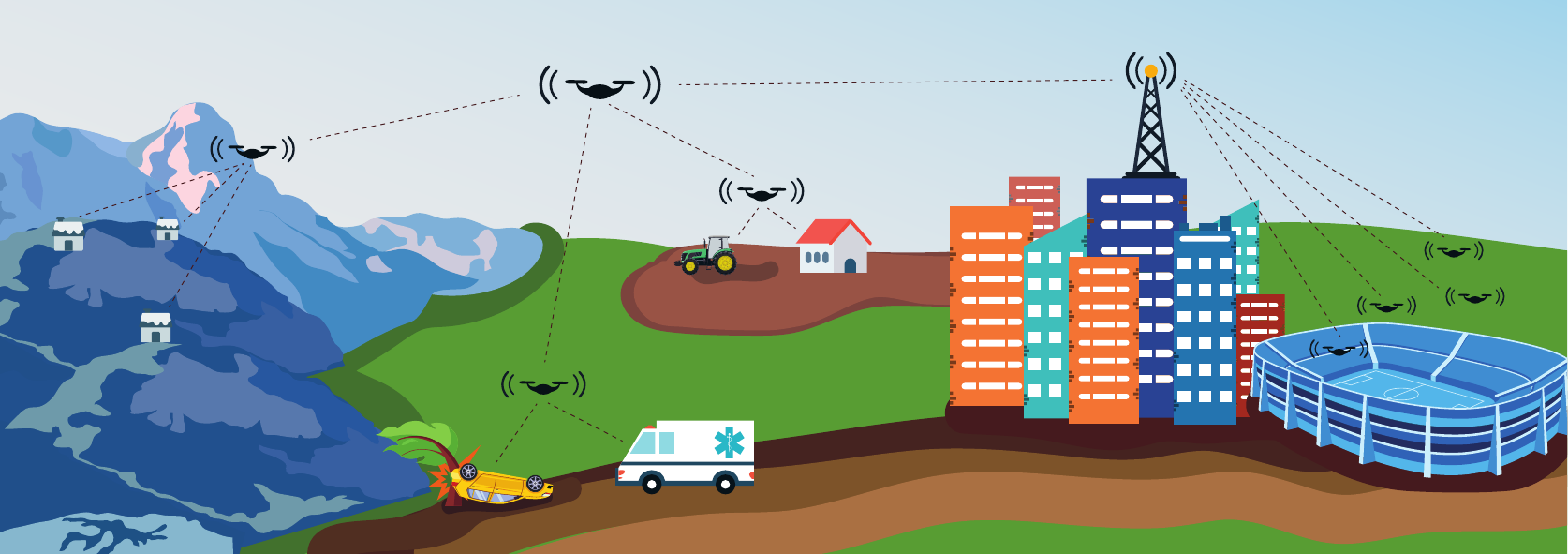}
 \caption{A network of aerial base stations extends cellular coverage to overloaded areas or regions without infrastructure. }
 \label{fig:setup}
\end{figure}

\cmt{Contribution1}The goal of this paper is to address the
aforementioned limitations. To the best of our knowledge, \emph{it is
the first work to propose a framework for  
 multiple-AirBS placement in a fully adaptive and decentralized
fashion, without any need for communication or coordination among
UAVs.} Besides its  simplicity, the proposed
scheme features low computational and communication 
requirements. The technical approach involves two steps. 
\begin{myitemize}%
\myitem\cmt{description}%
\begin{myitemize}%
\myitem\cmt{utility}%
\begin{myitemize}%
  \myitem\cmt{description}First, a network utility is designed so that
  each AirBS can determine the improve directions without information from
  other AirBSs. 
 \myitem\cmt{benefits}This naturally renders the
  developed algorithms non-cooperative, decentralized, and robust
to   communication failures.
\myitem\cmt{smoothness}A key idea is
to leverage smooth surrogate functions from machine learning to
construct a continuous and differentiable objective. This property is
critical for adaptive implementations, since  non-smooth criteria
(e.g. mixed-integer programs, \cite{kim2018topology}) may generate
oscillations and erratic behaviour 
with  slightly perturbed inputs. 

\end{myitemize}%
\myitem\cmt{stochastic opt.}
\begin{myitemize}%
\myitem\cmt{description}Second, the designed utility is maximized
invoking tools from stochastic optimization, which constitutes the
workhorse of deep neural network training given their simplicity and
low computational requirements. 
\myitem\cmt{benefits}%
\begin{myitemize}%
\myitem\cmt{lightweight}The resulting placement algorithm inherits
these lightweight features,
\myitem\cmt{convergence}convergence guarantees,
\myitem\cmt{adaptive}and can adapt to changes in an \emph{online}
fashion. This includes changes in  MU data usage,
MU location, or even the number of MUs or AirBSs. 
\myitem\cmt{novelty}Together with the aforementioned
smooth surrogate functions, applying stochastic optimization
constitutes the main novelty of this 
work.\footnote{Although
  \cite{andryeyev2016selforganized} claims to perform ``stochastic
  gradient search'' for AirBS placement
  \cite[Fig. 2(b)]{andryeyev2016selforganized}, their approach is not
  related 
  to stochatic optimization. Instead, artificial noise is added to the
  trajectory to avoid local 
  optima.
}
\end{myitemize}%
\end{myitemize}%
\end{myitemize}%
\end{myitemize}%


\cmt{paper structure}The rest of the paper is structured as
follows. Sec.~\ref{sec:model} describes the model and placement
problem. Sec.~\ref{eq:placement} proposes a framework for
multiple-AirBS placement based on stochastic optimization. 
Finally, Sec.~\ref{sec:sim} validates the proposed scheme through
simulations  and Sec.~\ref{sec:conclusions} summarizes 
conclusions.

\section{Model and Goal}
\label{sec:model}

\cmt{overview}Consider a setup where $\bsnum$ AirBSs, each one
assembled on board an autonomous rotorcraft\footnote{Examples of
  rotorcraft, also called multicopters, include quadcopters and
hexacopters. The proposed scheme cannot directly accommodate
fixed-wing UAVs since they are unable to hover at a fixed
location. }, must provide connectivity between a collection of
$\munum$ MUs and the terrestrial cellular
infrastructure. \nextv{Although the MUs may be thought to be on the ground,
the same scheme applies if the MUs are other UAVs, which may
be carrying out e.g. a surveillance or video recording
mission.} \cmt{downlink}For clarity, the discussion focuses on
the downlink; yet the scheme  
can  readily  accommodate  the uplink.  Each MU is
 associated with a single AirBS, which 
receives data packets from the terrestrial infrastructure an sends them
to the MU.\cmt{backhaul} This process may also be performed in multiple
hops, so a packet is relayed by multiple AirBSs before reaching the
MU. The (possibly multi-hop) link between the infrastructure and the
serving AirBS will be referred to as backhaul. Throughout, it will be
assumed that the AirBSs
can establish a backhaul connection with the terrestrial infrastructure
in the entire geographic area of interest. This is a reasonable
assumption unless the target area is too large for the number and
range of the AirBSs, which would require  additional
considerations. \nextv{Indeed,  it
is even assumed in some works that the backhaul takes place through a
satellite 
link. }
\cmt{layer}The
AirBSs may operate as relays at the physical layer or even 
have  upper-layer capabilities like  eNodeBs in LTE.

\begin{myitemize}%
\myitem\cmt{AirBSs}
\begin{myitemize}%
\myitem\cmt{location}The location of the $\bsind$-th AirBS is represented by 
its vector $\bslocvec\bsnot{\bsind}\in\rfield^3$ of spatial coordinates. 
\end{myitemize}%
\myitem\cmt{Mobile Users (MUs)}%
\begin{myitemize}%
\myitem Similarly, the location of the $\muind$-th MU is given by 
$\mulocvec\munot{\muind}\in \rfield^3$. 
\end{myitemize}%
\myitem\cmt{Channel}%
\begin{myitemize}%
  \myitem\cmt{def}The downlink channel (i.e. the data channel from the
  AirBSs to the MUs) is characterized by a function
  $\bslocvec\bsnot{\bsind}\mapsto
  \chgain\bsmunot{\bsind}{\muind}(\bslocvec\bsnot{\bsind},\mulocvec\munot{\muind})$
  which provides the channel gain between the the
  $\bsind$-th AirBS and the  $\muind$-th MU when they respectively lie at locations
$\bslocvec\bsnot{\bsind}$ and  $\mulocvec\munot{\muind}$.  Clearly,
  this quantity is determined by the antenna gains
  and propagation phenomena.  To simplify the notation, let
  $\txpower\bsnot{\bsind}$ be the power transmitted by the $\bsind$-th
  AirBS and let
  $
  \rxpower\bsmunot{\bsind}{\muind}(\bslocvec\bsnot{\bsind},\mulocvec\munot{\muind})
  \define \txpower\bsnot{\bsind}
  \chgain\bsmunot{\bsind}{\muind}(\bslocvec\bsnot{\bsind},\mulocvec\munot{\muind})
  $ denote the power received by the $\muind$-th MU from the
  $\bsind$-th AirBS.
\myitem\cmt{knowledge}The $\bsind$-th AirBS is assumed to know the
gradient of
$ \rxpower\bsmunot{\bsind}{\muind}(\bslocvec\bsnot{\bsind},\mulocvec\munot{\muind})$
with respect to $\bslocvec\bsnot{\bsind}$. In practice, it is not possible to
know $
\rxpower\bsmunot{\bsind}{\muind}(\bslocvec\bsnot{\bsind},\mulocvec\munot{\muind})$
exactly, so a certain performance degradation is expected due to errors
in  this model. 

\myitem\cmt{example:free space}Although  the
  proposed scheme can accommodate any model
  $
  \chgain\bsmunot{\bsind}{\muind}(\bslocvec\bsnot{\bsind},\mulocvec\munot{\muind})$
  in the literature (see e.g.~\cite{zeng2019accessing} for a
  survey) so long
  as this function is differentiable with respect to
  $\bslocvec\bsnot{\bsind}$, a simple example 
is 
\begin{myitemize}%
  \myitem\cmt{description}free space propagation. This 
  choice is sometimes reasonable since air-ground channels have often a line of
  sight; see e.g.~\cite{zeng2019accessing}. In that case, $
\rxpower\bsmunot{\bsind}{\muind}(\bslocvec\bsnot{\bsind},\mulocvec\munot{\muind})$
is given by
\begin{align}
\label{eq:freespace}
  \rxpower\bsmunot{\bsind}{\muind}(\bslocvec\bsnot{\bsind},\mulocvec\munot{\muind}) = \frac{\propconst_{\bsind,\muind}
  \txpower\bsnot{\bsind}}{
  ||\bslocvec\bsnot{\bsind}-\mulocvec\munot{\muind}||^2},
  \end{align}
where  $\propconst_{\bsind,\muind}$ represents the channel gain \nextv{when
the $\bsind$-th AirBS and the $\muind$-th MU lie }at unit distance. It
depends on $\bsind$ and $\muind$ due to the influence of the antenna
patterns as well as the  low-noise and power amplifiers. 
\myitem\cmt{Limitation}Although the model \eqref{eq:freespace} is 
simple and tractable, it is not appropriate for height optimization;
see e.g.~\cite[Sec. III-C]{zeng2019accessing}.
Still, one  may adopt \eqref{eq:freespace} to
set the horizontal position, i.e. the first two entries of
$\bslocvec\bsnot{\bsind}$, and determine a constant suitable 
height  e.g. as in \cite{alhourani2014lap}.
\end{myitemize}%
\end{myitemize}%

\myitem\cmt{low-bandwidth control channel}
\begin{myitemize}%
\myitem\cmt{Motivation}Besides the backhaul and downlink channel, note
that there must exist a \emph{control} channel that allows communication between
the AirBSs and MUs even before the AirBSs have arrived at the
vecinity of the  MUs. This is necessary because the AirBSs need to know at least the
approximate locations of the MUs to navigate to a suitable position. 
\myitem\cmt{implementation}This can be implemented as a  a satellite or
low-frequency (and therefore long-range 
and low-rate) terrestrial channel. 
\myitem\cmt{assumptions}In this work, only minimum requirements will be imposed on this
control channel. Specifically, it is assumed that each MU can send
short \emph{control packages} through this channel and that all AirBSs
receive them. It is not even needed that this channel allows
bidirectional communication.
\end{myitemize}%
\end{myitemize}%

\cmt{broad goal}
\begin{myitemize}%
\myitem\cmt{placement}The problem of AirBS placement is that of selecting 
$\bslocvec\define[\bslocvec\bsnot{1}\transpose,\ldots,
\bslocvec\bsnot{\bsnum}\transpose ]\transpose$  to maximize a
given network utility that quantifies the QoS
experienced by the MUs.
\myitem\cmt{decentralized}The goal of this paper is to solve this
problem in an adaptive and decentralized
fashion without  cooperation or control communication among AirBSs.
Achieving this goal would increase safety and enable swift and
flexible deployments of AirBSs, possibly with inexpensive equipment. \opt{Note
that adaptability is not only critical to accommodate changes in the
channel, MU locations, or MU data requirements, but also to
accommodate changes in the number of AirBSs. The latter aspect is
important since these devices are typically powered by batteries which
need to be frequently recharged. }

\myitem\cmt{Non-convex}The problem of AirBS placement with
any reasonable network utility has multiple local optima and therefore
is intrinsically non-convex. To see
this, suppose that all AirBSs are identical and transmit with the same
power. Then, permuting the locations of the AirBSs arranged
in a locally optimal placement would yield another locally optimal
placement. There exist approaches intended to find
globally optimal placements (see e.g.~\cite{kim2018topology}), but
they require a central processor and \nextv{therefore suffer from the
limitations outlined in Sec.~\ref{sec:intro}. Furthermore, even such global
optimization techniques} are not even guaranteed to find  global optima. 
Therefore, the main concern should not be to reach
global optima but a reasonable local optimum; see
also~\cite[Sec.~III-D]{zeng2019accessing}.
\end{myitemize}%

\section{Adaptive and Non-cooperative Placement}
\label{eq:placement}

\cmt{Overview}As described in Sec.~\ref{sec:intro} and detailed in
this section, the proposed
framework comprises (i) a suitable network metric designed so that the
AirBSs can update their locations without cooperating and (ii) a
stochastic optimization algorithm to optimize such a metric. 

\cmt{Network metric}
\begin{myitemize}%
\myitem\cmt{uniform Utility function}Consider a network utility function of the form
\begin{align}
\label{eq:utility}
J(\bslocvec) = \frac{1}{\munum}\sum_{\muind=1}^\munum J_\muind(\bslocvec),
\end{align}
where $J_\muind(\bslocvec) $  quantifies the QoS experienced by the
$\muind$-th MU.  For instance, 
$J_\muind(\bslocvec)$  may  be given by
 (see also Sec.~\ref{sec:utility} for more details on
this and other functions)
\begin{align}
\label{eq:junicast}
  J_\muind(\bslocvec) = \log_2\left(
1 +  \frac{\max_\bsind \rxpower\bsmunot{\bsind}{\muind}(\bslocvec\bsnot{\bsind},\mulocvec\munot{\muind})}{\noisepower}
\right), 
\end{align}
where $\noisepower$ is the noise power.

\myitem\cmt{user-dependent}If the operator wishes that the AirBSs
favor areas with heavier traffic demands, one may generalize the
average in \eqref{eq:utility} to assign a greater weight to those
users with higher data rates, as described next. Among all packets
received by the AirBS network from the terrestrial infrastructure to
be delivered to the MUs, let 
$\probpack\munot{\muind}\in[0,1]$ denote the fraction of those packets
that are intended for the $\muind$-th user,
$\muind\in\{1,\ldots,\munum\}$.  In that 
case, one could think of replacing \eqref{eq:utility} with
\begin{align}
\label{eq:weightedutility}
J(\bslocvec) =\sum_{\muind=1}^\munum \probpack\munot{\muind}
  J_\muind(\bslocvec), 
\end{align}
which therefore quantifies the average QoS per packet.\footnote{Moving
  from \eqref{eq:utility} to \eqref{eq:weightedutility} is not
  necessary to apply the proposed scheme, but it will be
  instructive to  understand the pursued
stochastic optimization approach.}
\nextv{\myitem\cmt{purely optional}It is important to emphasize that this extension is not
needed and the proposed scheme can readily assign the same weight to all users as in
\eqref{eq:utility} just by setting $\probpack\munot{\muind} =
1/\munum$. However, it is also instructive in order to understand the pursued
stochastic optimization approach.
}
\myitem\cmt{ensemble mean}Since $\sum_\muind \probpack\munot{\muind}=1$, one can equivalently
express \eqref{eq:weightedutility} as
\begin{align}
\label{eq:expectedutility}
J(\bslocvec) =\expected\left[ J_\muind(\bslocvec) \right],
\end{align}
where $\muind$ in \eqref{eq:expectedutility} is a random variable that
takes the value $\auxmuind\in\{1,\ldots,\munum\}$ with probability
$\probpack\munot{\auxmuind}$. Thus, $\muind$ can be thought of as a random variable that
indicates a  packet recipient and  $\probpack\munot{\muind}$ is the
probability that a given packet must go to the $\muind$-th MU. Note
that $\{\probpack\munot{\muind}\}_{\muind=1}^\munum$ are dictated by
the infrastructure and thus cannot be modified by the AirBSs. 

\myitem\cmt{unknown}Although AirBSs may know the functional form of
$J(\bslocvec)$, they may not be able to evaluate it since it depends on
unknown variables or parameters. For
example, the AirBSs may know that $J_\muind(\bslocvec)$ is of the form
\eqref{eq:junicast} but they will not generally know
$\max_\bsind
\rxpower\bsmunot{\bsind}{\muind}(\bslocvec\bsnot{\bsind},\mulocvec\munot{\muind})$
for \emph{all} MUs at \emph{all} times. Collecting this information,
which furthermore is subject to constant change, would certainly
require a complicated methodology and would be challenging to implement in
a decentralized fashion. As seen later, stochastic
optimization bypasses this difficulty and  allows the AirBSs to minimize $J(\bslocvec)$ without
even knowing the number of MUs or AirBSs in the~system.

\myitem\cmt{Intuition}But before delving into that, it is convenient
to develop some intuition. To this end, note that the utility $ J(\bslocvec)$ in
\eqref{eq:expectedutility} for each (fixed) value of $\bslocvec$ can
be estimated by considering $\packnum$ packets that the AirBS network
receives from the terrestrial infrastructure over a
certain time interval. Specifically, suppose that it receives
$\packnum$ packets and that the $\packind$-th packet has to be
delivered to the  $\muind[\packind]$-th MU through the downlink
of the associated AirBS.  Suppose also that, upon receiving the
corresponding packet,  the $\muind[\packind]$-th 
MU sends through the control channel the information that the AirBSs
need to calculate $J_{\muind[\packind]}(\bslocvec)$. For example, if 
$J_{\muind[\packind]}(\bslocvec)$ is given by \eqref{eq:junicast},
then the $\muind[\packind]$-th MU sends\footnote{Although the
  $\muind$-th MU may measure the power 
  $\rxpower\bsmunot{\bsind}{\muind}(\bslocvec\bsnot{\bsind},\mulocvec\munot{\muind})$
  of (potentially) all AirBSs using their beacons,  it
  is only associated with one of them. 
}
$\max_\bsind
\rxpower\bsmunot{\bsind}{\muind[\packind]}(\bslocvec\bsnot{\bsind},\mulocvec\munot{\muind[\packind]})$
(assume for simplicity that the AirBSs know $\noisepower$). With this
information, the AirBSs can obtain 
$\{  J_{\muind[\packind]}(\bslocvec)\}_{\packind=1}^\packnum$ and
therefore 
\begin{align}
\label{eq:expectedutilitysample}
\hat J(\bslocvec) =\frac{1}{\packnum} \sum_{\packind=1}^\packnum  J_{\muind[\packind]}(\bslocvec),
\end{align}
which is an unbiased estimator of
$J(\bslocvec)$ under general conditions. By the law of the large numbers,
$\hat J(\bslocvec) $ converges to $J(\bslocvec) $ with probability 1
as $\packnum\rightarrow \infty$ if the indices
$\{\muind[\packind]\}_\packind$ are independent or if they make up an ergodic stochastic process. 

The bottomline is that the AirBSs can \nextv{use
  \eqref{eq:expectedutilitysample} to} estimate $J(\bslocvec)$ by just
receiving information from a small fraction of MUs, which is more
practical than maintaining a real-time database per AirBS with the information from all
MUs. Although the scheme in the next section
does not estimate $J(\bslocvec)$ but its gradient, the underlying idea
is the same as illustrated with this toy example.



\myitem\cmt{one sample}Finally, note that
\eqref{eq:expectedutilitysample} provides a valid estimator for
$J(\bslocvec)$ even if $\packnum=1$, yet in this case the estimates
will be substantially noisy. Stochastic algorithms, like the one in
Sec.~\ref{sec:navigator},  implicitly
introduce averaging to counteract this effect.

\end{myitemize}%

\subsection{Adaptive Stochastic Navigator}
\label{sec:navigator}

\cmt{Overview}As a step towards the targeted technology, this section
describes a technique that enables AirBS to 
update their location with only information that can be easily collected
 in practice and from only a few MUs at each time.  Sec.~\ref{sec:utility} will design utilities
$J(\bslocvec)$ that allow  location updates without
information on the other AirBSs.

\cmt{centralized solution}
\begin{myitemize}%
\myitem\cmt{description}If there were a central controller with
real-time access to all relevant system information and ideal
communication links to all AirBSs, then $J(\bslocvec)$ could be
maximized  e.g. via gradient ascent as
\begin{align}
\label{eq:centralizedgradientascent}
\bslocvec\itnot{\itind+1} = \bslocvec\itnot{\itind} 
+ \stepsize\itnot\itind \nabla J (\bslocvec\itnot{\itind} ) ,
\end{align}
where $\itind=0,1,\ldots$ is the iteration index,
$\stepsize\itnot\itind>0$ is a step size, 
 $\bslocvec\itnot{0}$
is the  initial placement, and (cf. \eqref{eq:weightedutility}) 
\begin{align}
\label{eq:centralizedgradient}
\nabla J(\bslocvec) =  \sum_{\muind=1}^\munum \probpack\munot{\muind}\nabla J_\muind(\bslocvec) .
\end{align}
\cmt{no constraints assumption}Constraints on $\bslocvec$
could also be accommodated e.g. to impose a minimum safety distance between
AirBSs, but this possibility is disregarded here to simplify the
exposition.  \myitem\cmt{limitations}Unfortunately, the centralized approach in
\eqref{eq:centralizedgradientascent} is problematic in practice.
\begin{myitemize}%
  \myitem\cmt{comms}First, failures in the communication links between
  the AirBSs and the central controller would limit the capacity of
  AirBSs to navigate to appropriate locations and could even compromise
  operational safety.  \myitem\cmt{need for info}Second, evaluating
  $\nabla J_\muind(\bslocvec) $ would generally require information on all MUs and
  AirBSs such as the communication channel between all MUs and all AirBSs,
  their locations and so on; see also the discussion earlier in 
  Sec.~\ref{eq:placement}. But
  expecting such a hypothetical central controller to gather this information in real time is generally unrealistic.
  \myitem\cmt{prob.}\opt{Besides,  evaluating
  \eqref{eq:centralizedgradient} would also require estimates of the (possibly time-varying)
  probabilities $\{\probpack\munot{\muind}\}_\muind$, which entails additional overhead.}
\end{myitemize}%

\end{myitemize}%


\cmt{stochastic gradient ascent}
\begin{myitemize}%
\myitem\cmt{idea}A key idea in the proposed
framework is to sidestep these difficulties by capitalizing on stochastic
optimization methods. These methods stem from the observation that 
$\nabla J(\bslocvec)$ in \eqref{eq:centralizedgradient} can be
expressed as $\nabla J(\bslocvec) = \expected\{\nabla
J_\muind(\bslocvec)\}$ and replaced with an estimate, as done for
$J(\bslocvec)$ in
\eqref{eq:expectedutilitysample}. 
The idea is to
update $\bslocvec$ every time an MU (or a certain number of MUs) sends
the relevant information through the control channel. 
\myitem\cmt{update}Specifically, suppose that at time  $t_\itind$, the
AirBS network receives the $\itind$-th packet from the terrestrial
infrastructure  and that it must be delivered
to the  $\muind[\itind]$-th MU. Upon receiving this
packet, the    $\muind[\itind]$-th MU uses the control channel to send
 the information that the AirBSs need to compute 
$\nabla J_{\muind[\itind]}(\bslocvec)$. The AirBSs may then update
their positions through  a \emph{stochastic gradient
  } step:
\begin{align}
\label{eq:updateall}
\bslocvec\itnot{\itind+1} = \bslocvec\itnot{\itind} 
+ \stepsize\itnot\itind \nabla J_{\muind\itnot{\itind}}(\bslocvec\itnot{\itind} ).
\end{align}
Similarly to what was described around
\eqref{eq:expectedutilitysample}, $\nabla
J_{\muind\itnot{\itind}}(\bslocvec\itnot{\itind} )$ constitutes an
unbiased estimate of $\nabla J(\bslocvec\itnot{\itind})$. 
This is the same idea utilized by the classical \emph{least mean
  squares} (LMS) algorithm in signal processing. 

\myitem\cmt{minibatches}Unlike \eqref{eq:centralizedgradientascent},
which requires information from all MUs, the update
\eqref{eq:updateall} only involves information from the
$\muind[\itind]$-th MU. The caveat is that the gradient estimates
$\nabla J_{\muind\itnot{\itind}}(\bslocvec\itnot{\itind} )$ are
\emph{noisy}. To alleviate this effect, it is customary in stochastic
optimization to average several of these gradient estimates before performing
each update. In this case, this means that the AirBSs may update their
position only every $\minibatchnum$  packets, where $\minibatchnum$
is referred to as the \emph{minibatch} size.



\end{myitemize}%

\cmt{strengths of stoch. opt}Clearly, the stochastic update in
\eqref{eq:updateall} constitutes a valuable alternative to
\eqref{eq:centralizedgradientascent}. 
\begin{myitemize}%
\myitem\cmt{widely used}Since stochastic gradient methods enjoy high
popularity, 
\myitem\cmt{convergence rate}their convergence is well analyzed. Due to space limitations,
we omit  those results here, but they can be found 
e.g. in~\cite{bottou2018large}.
\myitem\cmt{no memory}Note also that almost no memory is required
\myitem\cmt{adapt to changes}and, in part for this reason, the update can adapt
to system changes.
\end{myitemize}%


\textit{Remark 1.}\cmt{dynamics}
\begin{myitemize}
\myitem\cmt{stepsize}The step size
$\stepsize\itnot\itind$ must be chosen in accordance with the dynamic restrictions
of the UAVs, such as their maximum horizontal velocity. 
\myitem\cmt{waypoints}The sequence $\{\bslocvec\bsnot{\bsind}\itnot{\itind}\}_\itind$
may be interpreted as a sequence of waypoints. The autopilot of each
UAV will then issue low-level control commands to the rotors to follow
such a sequence. Because of their dynamics, the UAVs are not capable of
accurately following arbitrary waypoint sequences and, hence, the actual
trajectory may be a smoothed or ``filtered'' version of the one
indicated by the waypoints. Beforehand, this need not be a limitation
since the  gradient estimates, and hence
$\{\bslocvec\itnot\itind\}_\itind$,  are 
intrinsically noisy. The aforementioned smoothing effect may
even be beneficial for maximizing $J(\bslocvec)$; see~\cite{bottou2018large}.
\end{myitemize}%

\subsection{Utility Functions for Non-cooperative Placement}
\label{sec:utility}

\begin{figure}[t]
 \centering
 \includegraphics[width=0.45\textwidth]{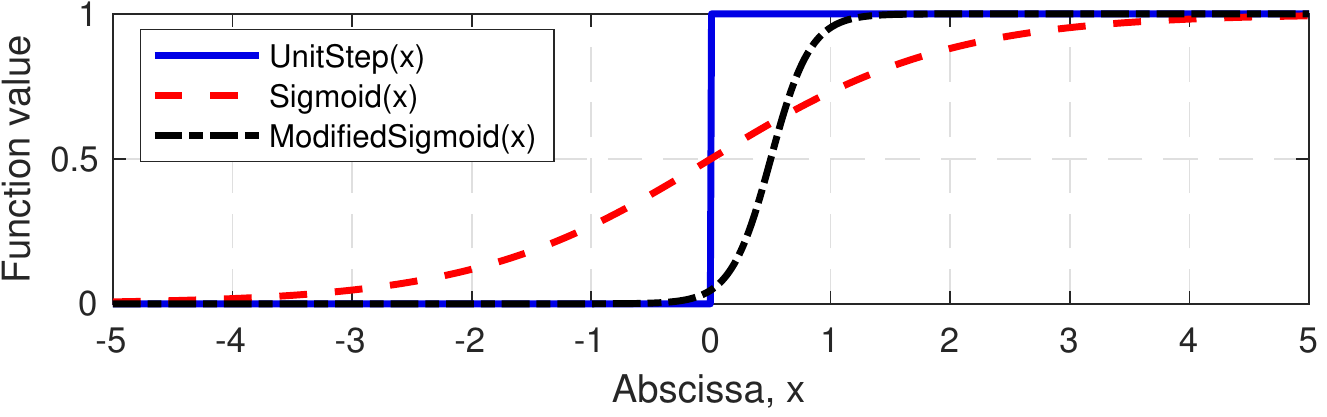}
 \caption{The proposed modified sigmoid function will be used as a
   differentiable surrogate of the unit step ($\Delta = 1$). }
 \label{fig:modsigmoid}
\end{figure}

\cmt{Update}Equation
\eqref{eq:updateall} provides the update for all AirBS locations. It implies that the
$\bsind$-th AirBS must update its position as
\begin{align}
\label{eq:updateone}
\bslocvec\bsnot{\bsind}\itnot{\itind+1} = \bslocvec\bsnot{\bsind}\itnot{\itind} 
+ \stepsize\itnot\itind
\nabla_{\bslocvec\bsnot{\bsind}}J_{\muind\itnot{\itind}}(\bslocvec\itnot{\itind}).
\end{align}
\cmt{Objective design}To apply this  scheme
without  cooperation
or communication among AirBSs, the user utility $J_\muind$ must be
chosen so that each AirBS can compute
$\nabla_{\bslocvec\bsnot{\bsind}}J_\muind(\bslocvec\itnot{\itind})$
without the need for information from other AirBSs. 
\cmt{function of powers}To this end, the key idea here is to   focus  on functions
$J_\muind(\bslocvec) $ that can be expressed as
\begin{align}
\label{eq:thrup}
J_\muind(\bslocvec) = f(
\rxpower\bsmunot{1}{\muind}(\bslocvec\bsnot{1},\mulocvec\munot{\muind})
,\ldots,
\rxpower\bsmunot{\bsnum}{\muind}(\bslocvec\bsnot{\bsnum},\mulocvec\munot{\muind})
)
\end{align}
for some $f:\rfield^\bsnum\rightarrow \rfield$. This is not a highly
restrictive requirement since many usual network utilities, such as the
sum rate, are indeed of this form. To see that such functions achieve
this goal, note from the chain rule that
\begin{align}
\nonumber
&\nabla_{\bslocvec\bsnot{\bsind}}J_{\muind\itnot{\itind}}(\bslocvec\itnot{\itind}) = 
\left[\nabla_{\bslocvec\bsnot{\bsind}}\rxpower\bsmunot{\bsind}{\muind\itnot{\itind}}(\bslocvec\bsnot{\bsind}\itnot{\itind},\mulocvec\munot{\muind\itnot{\itind}})
\right] \cdot 
\\\label{eq:gradientprod}
&~~~\bigg[\frac{\partial}{\partial z}f(
\rxpower\bsmunot{1}{\muind\itnot{\itind}}(\bslocvec\bsnot{1}\itnot{\itind},\mulocvec\munot{\muind\itnot{\itind}})
,\ldots, z,\ldots,\\\nonumber&\hspace{2cm}
\rxpower\bsmunot{\bsnum}{\muind\itnot{\itind}}(\bslocvec\bsnot{\bsnum}\itnot{\itind},\mulocvec\munot{\muind\itnot{\itind}})
)\bigg]_{ z =\rxpower\bsmunot{\bsind}{\muind\itnot{\itind}}(\bslocvec\bsnot{\bsind}\itnot{\itind},\mulocvec\munot{\muind\itnot{\itind}})},
\end{align}
where the dummy variable $z$ occupies the $\bsind$-th argument of
$f$. Thus,  the $\bsind$-th AirBS can obtain
$\nabla_{\bslocvec\bsnot{\bsind}}J_{\muind\itnot{\itind}}(\bslocvec\itnot{\itind})$ if it knows both
terms in brackets. 
\begin{myitemize}%
  \myitem{}The first can be obtained if the $\muind\itnot{\itind}$-th MU reports
  its location $\mulocvec\munot{\muind\itnot{\itind}}$ through the
  control channel  since
  the $\bsind$-th AirBS already knows its own location $\bslocvec\bsnot{\bsind}\itnot{\itind}$  and
  the gradient of
  $\rxpower\bsmunot{\bsind}{\muind}(\bslocvec\bsnot{\bsind},\mulocvec\munot{\muind})$;
  cf. Sec.~\ref{sec:model}. 
  \myitem The second term in brackets in \eqref{eq:gradientprod} can be computed by the
  $\muind\itnot{\itind}$-th MU and  sent likewise  to the  AirBSs through the
  control channel, since it only needs to measure the power
  received from the AirBSs. This can be done using e.g. their beacons. 
\end{myitemize}%
\cmt{Sum up}To sum up, the $\muind\itnot{\itind}$-th MU sends its own
location and the second term in brackets through the control channel. With this information, the AirBSs 
estimate the gradient, which points in a direction of increasing
network utility $J(\bslocvec)$ on average.


\cmt{sum rate}It remains to design suitable functions
$J_\muind(\bslocvec)$ of the form \eqref{eq:thrup}. 
The most direct choice of $J_\muind(\bslocvec)$
is the rate of the $\muind$-th user, which in turn means that
$J(\bslocvec)$ is the expected rate the MUs. To obtain this rate,
one may consider two scenarios:
\begin{itemize}
\item (S1) Each MU is associated with the AirBS from which it receives
  most power. Since the area where the AirBSs need to be deployed is
  typically remote and therefore most of the cellular spectrum is
  empty, it is reasonable to assume that each AirBS operates in a
  different band and therefore there exists no inter-AirBS
  interference. This assumption may also be relaxed, but it will be
  adopted here for simplicity. Under these circumstances, the rate of the
  $\muind$-th MU is proportional to 
\begin{align}
\label{eq:rateunicast}
  \rate_\muind(\bslocvec) = \log_2\left(
1 +  \frac{\max_\bsind \rxpower\bsmunot{\bsind}{\muind}(\bslocvec\bsnot{\bsind},\mulocvec\munot{\muind})}{\noisepower}
\right).
\end{align}


\item (S2) The AirBSs simply relay the signal transmitted by the
  terrestrial infrastructure. No association is required. To some extent, these relays act as active reflectors. This may be of interest e.g. for
  broadcasting applications  as in sport events. Assuming no
  carrier phase synchronization among AirBSs\nextv{ and/or considering that
  the UAVs randomly hover around a waypoint}, the signals relayed by
  the AirBSs add at the $\muind$-th MU in an incoherent fashion, which
  yields a total received power of
  $\sum_{\bsind=1}^\bsnum
  \rxpower\bsmunot{\bsind}{\muind}(\bslocvec\bsnot{\bsind},\mulocvec\munot{\muind})$. The
  data rate is therefore proportional to 
\begin{align}
\label{eq:ratebroadcast}
\rate_\muind(\bslocvec) = \log_2\left(
1 +  \frac{\sum_{\bsind=1}^\bsnum \rxpower\bsmunot{\bsind}{\muind}(\bslocvec\bsnot{\bsind},\mulocvec\munot{\muind})}{\noisepower}
\right).
\end{align}
\end{itemize}

\cmt{fairness \ra step function}Thus, one can directly set
$J_\muind(\bslocvec)=\rate_\muind(\bslocvec)$, where
$\rate_\muind(\bslocvec)$ is given either by \eqref{eq:rateunicast} or
\eqref{eq:ratebroadcast}. However, it is well-known (see e.g. ~\cite{boryaliniz2016placement}) that
the sum rate is typically a poor network metric in terms of fairness
since the resulting AirBS placement may yield a high rate for a small
subset of MUs to the detriment of the rest of MUs, which may suffer
from a low rate. Thus, both in S1 and S2, it may be preferable to
pursue AirBS placements where a certain degree of fairness is
promoted. 
\begin{myitemize}%
\myitem\cmt{min rate}The solution proposed here is to assign a  utility of 1
to those users whose rate exceeds a pre-selected nominal value
$\minrate$ and 0 otherwise. A related idea has also been used in the single-AirBS scheme~\cite{boryaliniz2016placement}. 
This can be implemented by setting 
$J_\muind(\bslocvec)=\unitstep(\rate_\muind(\bslocvec)-\minrate)$,
where $\unitstep(\cdot)$ is the unit-step function, returning 1 for
positive arguments and 0 otherwise. 
\myitem\cmt{min power}Equivalently, one can impose a minimum
requirement on the SNR or, directly, a minimum $\minpower$ on the received power. This reads
as
$J_\muind(\bslocvec)=\unitstep(\maxpower_\muind(\bslocvec)-\minpower)$
for (S1) and 
$J_\muind(\bslocvec)=\unitstep(\sumpower_\muind(\bslocvec)-\minpower)$
for (S2), where $\maxpower_\muind(\bslocvec)\define \max_\bsind
\rxpower\bsmunot{\bsind}{\muind}(\bslocvec\bsnot{\bsind},\mulocvec\munot{\muind})$
and $\sumpower_\muind(\bslocvec)\define \sum_{\bsind=1}^\bsnum
\rxpower\bsmunot{\bsind}{\muind}(\bslocvec\bsnot{\bsind},\mulocvec\munot{\muind})$. 
\end{myitemize}%

\cmt{function surrogates}
\begin{myitemize}%
\myitem\cmt{replace unit step with sigmoid}
\begin{myitemize}%
\myitem\cmt{limitations unit step}Although adopting this step
function yields a network metric that promotes fairness, two
difficulties arise. First, $\unitstep(\cdot)$ is not differentiable.
Second, even if this non-differentiability is somehow fixed, the
resulting functions $J_\muind(\bslocvec)$ are flat for almost
all\footnote{That is, except for a set with Lebesgue measure 0.} values
of $\bslocvec$. This means that the gradient is zero in those points
and the update \eqref{eq:updateone} would yield no movement of the
AirBSs unless for very specific values of $\bslocvec\itnot{\itind}$.
\myitem\cmt{sigmoid}Drawing inspiration
  from the machine learning literature, a solution proposed here is  to replace
  $\unitstep(\cdot)$ with an appropriately modified sigmoid function. 
\begin{myitemize}%
  \myitem\cmt{def}The well-known sigmoid function is given by
  $\sigma(x) = e^x/(1+e^x)$ and  illustrated in
  Fig.~\ref{fig:modsigmoid}. Roughly speaking, it is close to zero for
  $x<-3$ and close to 1 for $x>3$. Therefore,
  $\sigma_\Delta(x)\define \sigma((6x/\Delta) - 3)$ exhibits the same
  transition but between $x=0$ and $x=\Delta$, where $\Delta$ is
  selected by the user. \myitem\cmt{gradient\ra is not zero}Note that its
  derivative is
  $\sigma'_\Delta(x) = (6/\Delta)\sigma'((6x/\Delta) - 3)$, where
  $\sigma'(x) = \sigma(x)(1-\sigma(x))$ is the derivative of
  $\sigma(x)$. Besides being differentiable everywhere,
  $\sigma'_\Delta(x)>0$ for all $x$ and therefore the iteration in
  \eqref{eq:updateone} will not stall unless the AirBSs
  are already in a locally optimal placement.
\end{myitemize}%
\end{myitemize}%
\myitem\cmt{non-differentiability of max}Additionally, the $\max$
operator in $\maxpower_\muind(\bslocvec)$  is another source for
non-differentiability and flat regions. Drawing inspiration from
deep learning, one can  replace this function by the log-sum-exp
function
$\logsumexpfun(\rxpower\bsnot{1},\ldots,\rxpower\bsnot{\bsnum})\define
\log(\sum_{\bsind}\exp\{\rxpower\bsnot{\bsind}\})$, whose gradient is
the well-known \emph{soft-max function}\nextv{, also popular in that area}. 

\end{myitemize}%

\section{Simulation Study}
\label{sec:sim}

\begin{figure}[t]
 \centering
\includegraphics[width=0.35\textwidth]{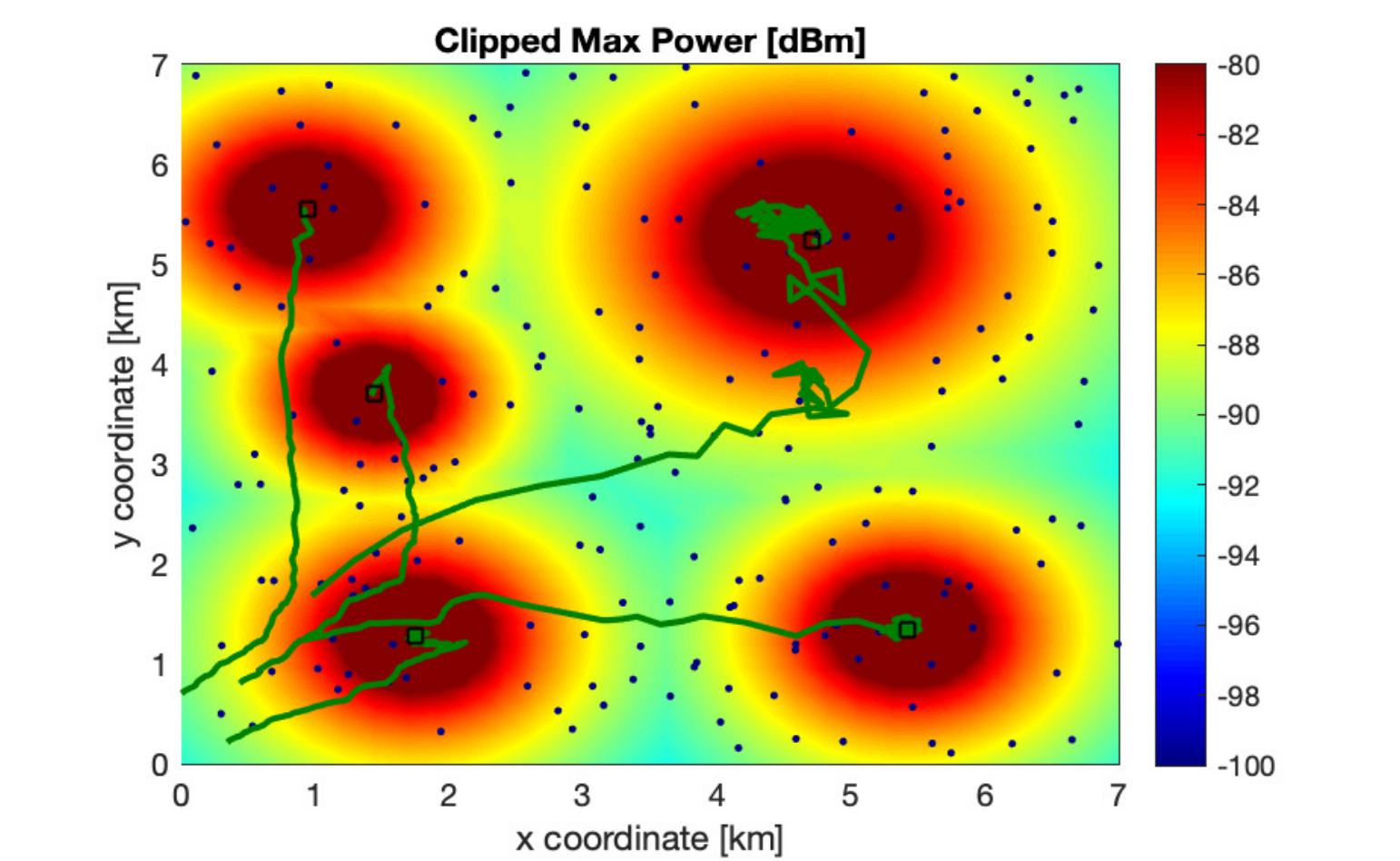}
 \caption{Trajectory followed by the AirBs (solid green lines). Squares
   indicate the position of the AirBSs after 100 updates. Dots indicate
   the positions of the MUs. The background color indicates the
   maximum of the power received from all AirBSs at each spatial
   location. It is observed that most MUs receive a power greater than
   the target value
 $\minpower=-91$ dBm.}
 \label{fig:path}
\end{figure}

To complement the theoretical convergence guarantees inherited from
stochastic gradient methods~\cite{bottou2018large}, this section
validates the  proposed scheme in a setup where  
$\bsnum=5$ AirBSs  act as picocells of an LTE system. 
\begin{myitemize}%
\myitem{}The main area  of interest is a square of $7\times7$ km. 
\myitem\cmt{base stations}
\begin{myitemize}%
\myitem{}The AirBSs are deployed initially uniformly at random
in the Southwest fourth of that area. 
\myitem{}Their transmitted power is given by
$[\txpower\bsnot{1},\ldots,\txpower\bsnot{\bsnum}]
= [7,9,9,9f,12]$ dBm per physical resource block (PRB).
\myitem{}The AirBSs update their positions 
\myitem{}via \eqref{eq:updateone} for $\itind=1,2,\ldots,100$
\myitem{}with a constant step size  $\stepsize\itnot\itind =
5$, $\forall \itind$, and a 
\myitem{}minibatch size $\minibatchnum=50$.
\myitem{}The downlink occupies a 20 MHz band at 2.385 MHz
(S-band).
\myitem{}AirBSs are equipped with an antenna that radiates only downwards with a
gain of 6 dBi. 
\myitem{}These parameters imply that the channel gain at 1 km from the
AirBS is approximately $\propconst_{\bsind,\muind}=-94$ dB $\forall
\bsind,\muind$ assuming that the MUs are equipped with a single
isotropic antenna and adopting the channel model \eqref{eq:freespace}.
\myitem{}Since this  model does not allow 
height optimization, the  height of the AirBSs is kept fixed to $\height =
30$ m; see Sec.~\ref{sec:model} for details and
alternatives.
\end{myitemize}%

\myitem\cmt{mobile users}
\begin{myitemize}%
\myitem{}A total of $\munum=200$ MUs are deployed uniformly at random across the
area. Two more users are respectively deployed out of the main
area of interest at positions $[35,35]$ km and
$[-35,35]$ km, where the origin is in the bottom left part of the
figures. 
\myitem{}The QoS they receive is quantified through \eqref{eq:rateunicast}.
\myitem\cmt{$\minpower$}To determine $\minpower$,
\begin{myitemize}%
\myitem{}let the noise power be -112.4 dBm per PRB, which is a typical
value in LTE~\cite[Clause 5.2.1.2]{3GPP36888}.
\myitem{}The goal is to attain  an SNR of 21.4 dB, which yields
90 \% of the maximum throughput of a
transport block size (TBS) of 84760 bits~\cite{3gpp2018modulation}.
This yields $0.9\cdot 84760
\text{bits}/1\text{ ms}\approx$ 762 kbps for every PRB. 
\myitem{}Thus, the minimum received power is set to $\minpower=
-112.4\text{ dBm } +21.4\text{ dB } = -91$ dBm.
\end{myitemize}%
\myitem{}The modified sigmoid parameter $\Delta$ is such that $\minpower+\Delta=-89$ dBm. 
\end{myitemize}%
\end{myitemize}%

\begin{figure}[t]
 \centering
 \includegraphics[width=0.35\textwidth]{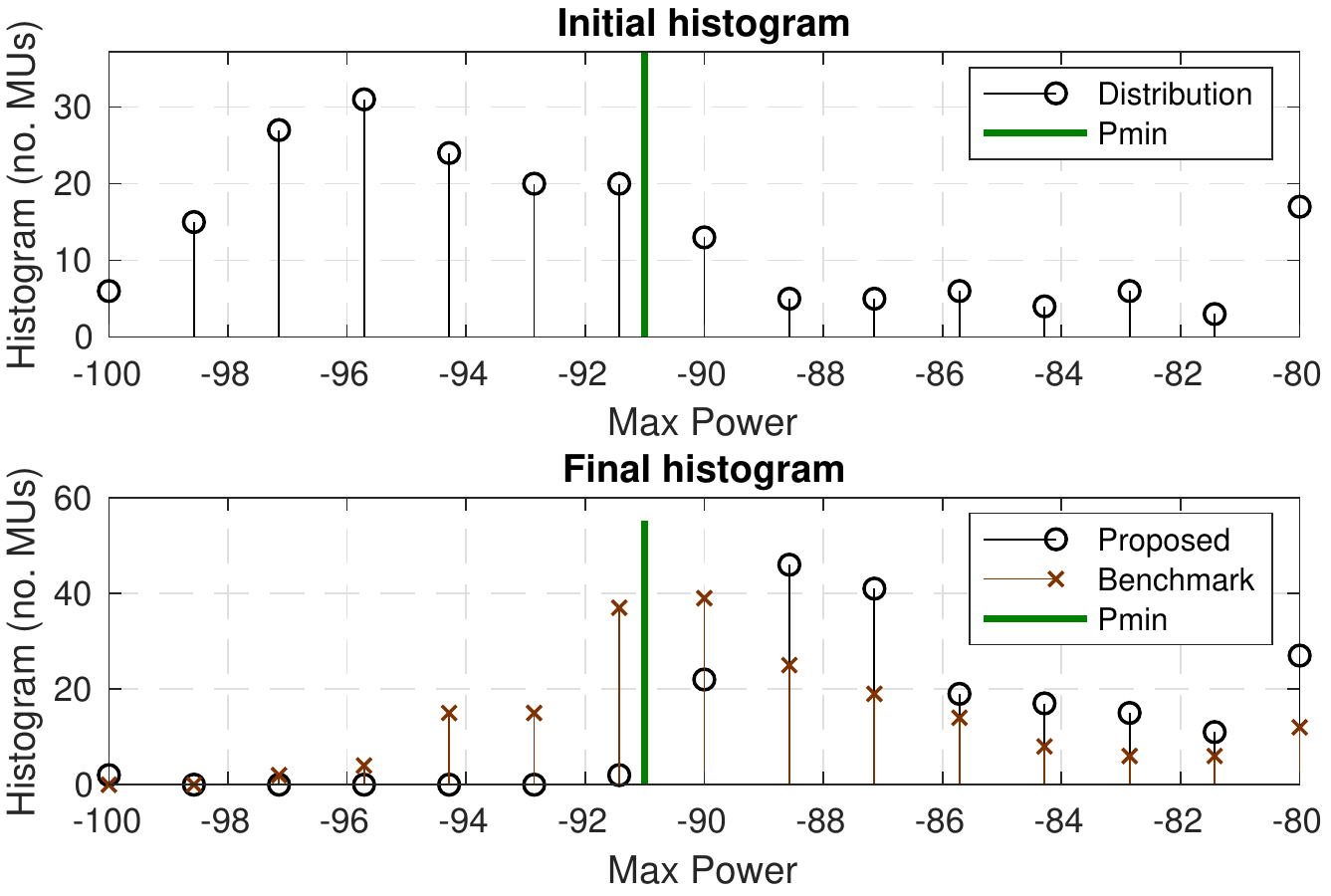}
 \caption{Histograms of the power of the strongest AirBS at each
   MU. At the beginning (top), the AirBSs are in their initial
   positions and have not found a suitable placement. After 100
   updates, the AirBSs adopt a position such that nearly all MUs
   receive a power greater than
   the target value
 $\minpower=-91$ dBm.}
 \label{fig:histograms}
\end{figure}

\cmt{Interpret figures}
\begin{myitemize}%
\myitem\cmt{trajectory}Fig.~\ref{fig:path} depicts the locations of the
MUs (except for the two MUs outside the main area of interest) with dots and the final location of the AirBSs with squares. For
visualization purposes, the
background color at each point  $\mulocvec$  indicates the result
of clipping the maximum power
$\maxpower(\bslocvec[100],\mulocvec) \define \max_\bsind
\rxpower\bsnot{\bsind}(\bslocvec\bsnot{\bsind}\itnot{100},\mulocvec)$ to the interval
$[-100,-80] $ dBm, where
$\rxpower\bsnot{\bsind}(\bslocvec\bsnot{\bsind}\itnot{100},\mulocvec)$ is the power
received from the $\bsind$-th AirBS at location $\mulocvec$ when the
AirBSs are in their final placement $\bslocvec\itnot{100}$. It is
observed that most of the area receives
$\maxpower(\bslocvec[100],\mulocvec)$ above $\minpower$. 
The paths followed by
the AirBSs (solid green lines) show that the AirBSs
naturally spread throughout the region even though they do not 
cooperate or communicate among them. Although the paths are somewhat noisy
due to the stochastic nature of the update, note that they just
correspond to waypoints -- the actual trajectories will be smoother; see Sec.~\ref{sec:navigator}. Observe that
the final arrangement accounts for the different transmit power of the
AirBSs. 

\myitem\cmt{histogram}Fig.~\ref{fig:histograms} depicts the histograms
of 
$\maxpower(\bslocvec[0],\mulocvec)$ (i.e. before
applying the proposed algorithm)
and
 $\maxpower(\bslocvec[100],\mulocvec)$ (i.e. after
 applying the algorithm). The final arrangement  meets  the
 target QoS  at 198 out of the 202 MUs. As a benchmark, the histogram
 is compared with 
 the one obtained if the AirBSs used K-means, which is the algorithm
 underlying the approaches in~\cite{liu2019deployment}
 and~\cite{hammouti2019mechanism}.  K-means performs poorly
 here because the two users off the area
 of interest shift the  centroids. In contrast,
 the  objective function designed in Sec.~\ref{sec:utility} allows the
 proposed algorithm to  ``give up'' those two remote users since
 serving them would require a placement $\bslocvec$ for which many of
 the users are not served. Indeed, the K-means algorithm fails to
 serve 73 users, which is 18 times more than the proposed
 method. Other algorithms in the literature are not fairly comparable
 with the proposed one since they require inter-AirBS communication or
 a central controller.

\end{myitemize}%

\cmt{more simulations}A video with more simulations can be found in
\cite{romero2019placementvideo}. The code will be posted on the
first author's website.

\section{Conclusions}
\label{sec:conclusions}

A framework has been developed for AirBS placement in a fully
non-cooperative, decentralized, and adaptive fashion. AirBSs move at
each iteration in a direction that improves the network utility on the
average. The
gradient of that utility is obtained via short messages
transmitted by the MUs through a low-bandwidth control
channel. Existing convergence analysis carries over  and performance is
validated in a simulation study. Future research 
will accommodate backhaul constraints for scenarios where the area of
interest is large relative to the number of AirBSs and their
communication range. 

\if\editmode1 
\onecolumn
\printbibliography
\else
\bibliography{\bibfilenames}
\fi
\end{document}

%% file: include.tex
\usepackage{fixltx2e}

\usepackage{graphicx}

\usepackage{subcaption}              

\usepackage[utf8]{inputenc}

\usepackage{amsfonts}
\usepackage{amsmath}
\usepackage{mathtools}

\usepackage{amssymb}

\usepackage{bm}

\usepackage{color,verbatim}
\usepackage{multirow}
\usepackage{accents}

\usepackage{theoremref}

\usepackage{url}

\newtheorem{myauxproblem}{Problem}

\newcounter{rulecounter}
\newcommand{\resetrule}{ \setcounter{rulecounter}{0}}
\resetrule

\newsavebox{\selvestebox}
\newenvironment{colbox}[1]
  {\newcommand\colboxcolor{#1}%
   \begin{lrbox}{\selvestebox}%
   \begin{minipage}{\dimexpr\columnwidth-2\fboxsep\relax}}
  {\end{minipage}\end{lrbox}%
   \begin{center}
   \colorbox{\colboxcolor}{\usebox{\selvestebox}}
   \end{center}}

\definecolor{orange}{rgb}{1,0.8,0}
\definecolor{gray}{rgb}{.9,0.9,0.9}
\definecolor{darkgray}{rgb}{.3,0.3,0.3}
\definecolor{darkblue}{rgb}{.1,0.0,0.3}
\definecolor{lightblue}{rgb}{0.7,0.7,1}
\definecolor{lightred}{rgb}{1,0.7,.7}
\definecolor{purple}{RGB}{204,153,255}
\definecolor{lightgray}{rgb}{.95,0.95,0.95}
\definecolor{lightgreen}{rgb}{0.6,0.8,0.6}
\definecolor{darkgreen}{rgb}{0.05,0.3,0.05}
\definecolor{pistachio}{RGB}{204,255,153}
\definecolor{paleturquoise}{RGB}{175,238,238}
\definecolor{yellow}{RGB}{255,255,153}

\newcommand{\ra}{$\rightarrow$~}

\newcommand{\bbm}[1]{{\bar{\bm #1}}}

\newcommand{\rfield}{\mathbb{R}}

\newcommand{\transpose}{^T}
 \newcommand{\define}{\triangleq}

\newcommand{\expected}[1]{\mathop{\textrm{E}}\brackets{#1} }

\newtheorem{myproposition}{Proposition}
\newtheorem{myquestion}{Question}
\newtheorem{myquiz}{Quiz}
\newtheorem{myremark}{Remark}
\newtheorem{myproblemstatement}{Problem Statement}
\newtheorem{mylemma}{Lemma}
\newtheorem{mytheorem}{Theorem}
\newtheorem{mydefinition}{Definition}
\newtheorem{mycorollary}{Corollary}
\newtheorem{myexample}{Example}